\documentclass[prl,aps,floatfix, preprint, showpacs]{revtex4}
\usepackage{amsmath,amssymb}
\usepackage{graphicx}

\begin{document}
%\draft

\title{Bounce-free Spherical Hydrodynamic Implosion}

\author{Grigory~Kagan, Xian-Zhu~Tang,
Scott~C.~Hsu and Thomas~J.~Awe}
\affiliation{
Los Alamos National Laboratory,
Los Alamos, NM 87545}

\date{\today}

\begin{abstract}
In a bounce-free spherical hydrodynamic implosion, the post-stagnation
hot core plasma does not expand against the imploding flow.  Such an implosion scheme has the advantage of
improving the dwell time of the burning fuel, resulting in a higher
fusion burn-up fraction.  The existence of
bounce-free spherical implosions is demonstrated by
explicitly constructing a family of self-similar solutions to the
spherically symmetric ideal hydrodynamic equations.  When applied to
a specific example of plasma liner driven magneto-inertial fusion, the bounce-free solution is found to
produce at least a factor of four improvement in dwell time and fusion energy gain.
\end{abstract}


\maketitle

Spherical hydrodynamic implosion is a central concept in inertial
confinement fusion (ICF)~\cite{Lindl} and magneto-inertial fusion
(MIF)~\cite{Lindemuth}, which use a variety of inertial pushers
driven by lasers~\cite{Atzeni}, heavy ion beams~\cite{heavy_ion},
and pulsed power~\cite{Intrator}.  It is also of fundamental
importance in late-stage stellar evolution, especially supernova core
collapse~\cite{Bethe}.  In a typical spherical hydrodynamic implosion,
converging mass flow provides compressional (and possibly shock)
heating of the core, converting flow kinetic energy into thermal
energy of the plasma localized about the center of symmetry. Once the
peak pressure is reached, a stagnation surface appears between the
core and imploding flow plasmas.  What happens next is usually a
bounce motion of the core in which the hotter plasma at the center
expands against the imploding flow.  In the special case where the
interface between the hot core and imploding flow remains stationary
rather than expanding outward, one has a bounce-free spherical
hydrodynamic implosion. That is, while in the typical bounce case both
the core/pusher interface and stagnation surface move radially
outward, in the bounce-free case the core/pusher interface remains
stationary and only the stagnation surface moves radially outward.  In
either case, 
once the stagnation surface sweeps through the
entire imploding material, the whole core-pusher assembly will undergo
an expansion.  The latter imposes an upper limit on the system dwell
time.

Optimizing toward a
bounce-free implosion provides the most
design freedom in increasing the dwell time and thereby a higher fusion burn-up fraction, which
as shown in this Letter, can be achieved by employing a specially shaped pusher. 
This approach is most naturally applied and would be of most benefit to plasma liner driven MIF~\cite{thio} since there
the pusher (liner) is created by merging plasma jets conceptually capable of modulating the imploding flow profile.
Also, due to the challenge for MIF to attain an ignited burn wave, it has the greatest need for increasing
the burn fraction via increased dwell time.  Conversely, the lack of a strong burn wave and intense hot spot pressures might actually make the
bounce-free implosion scheme energetically feasible to implement.

In this Letter we present a family of self-similar solutions to the
spherically symmetric ideal hydrodynamic equations which produce bounce-free
spherical implosions. The physics implications of these bounce-free
implosion solutions are explained in the application to
plasma liner driven MIF.  In particular, we find that by maintaining a bounce-free implosion through liner
(inertial pusher) profile shaping, the dwell time of the compressed
hot fuel can be improved by at least a factor of four compared with a
specific example of a stationary liner profile~\cite{parks-pop-2008},
while holding the liner speed and total energy the same. 

\begin{figure}[htbp]
\includegraphics[width=0.5\textwidth,keepaspectratio]{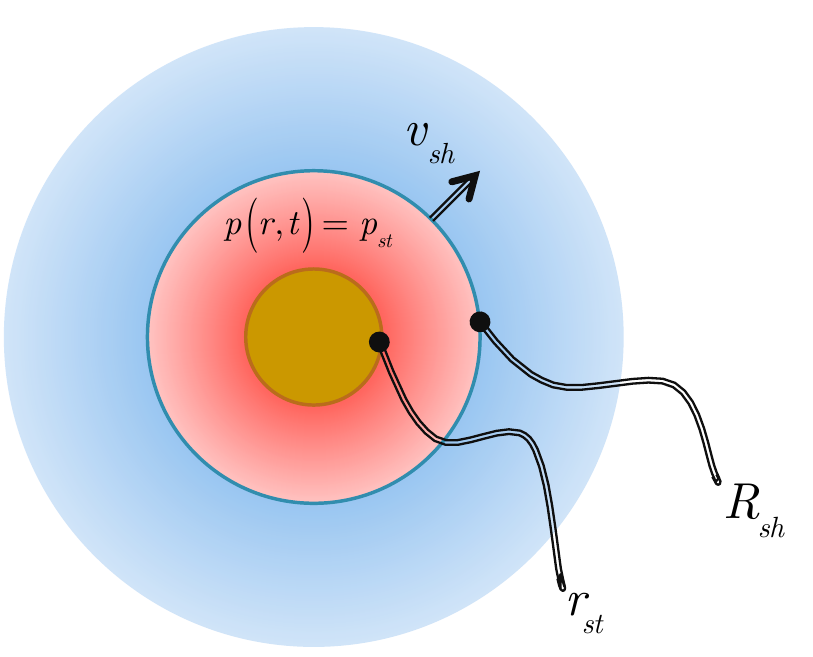}
\caption{Illustration of a bounce-free spherical
implosion. The target (in gold) stays still within its boundary located at
the stagnation radius $r_{\rm st}.$ The shocked part of the liner (in
red) is at rest with pressure and density being uniform and equal to $p_{\rm st}$ and
$\rho_{\rm st}$ respectively. The shocked front (denoted by $R_{\rm sh}$) is propagating outward, while the unshocked liner (in blue) is streaming  inward.}
\label{fig:sketch}
\end{figure}

To motivate the analytical solution, we first outline the physical
picture of and the mathematical constraints on bounce-free spherical
hydrodynamic implosions.  Figure~\ref{fig:sketch} shows a sketch of
the assembly after the liner hits the target. The stagnation or ignition pressure of the target $p_{\rm st}$
is achieved at time $t_{\rm st}$ with the spherical interface between the
hot spot and the liner at $r_{\rm st}$ coming to rest. To be bounce-free,
this interface remains still for $t>t_{\rm st}.$ This is
made possible by shock heating of the liner plasma neighboring $r_{\rm st}.$
In other words, a shock arises at the interface $r_{\rm
  st}$ at $t_{\rm st}$ and propagates radially outward. The shocked
liner in the region of $r_{\rm st} < r < R_{\rm sh}$ is in static
pressure balance with the target, $p(r,t)=p_{\rm st}.$ The shock front
at $R_{\rm sh}(t)$ becomes a second stagnation surface, which along with
the interface at $r_{\rm st}$, bounds an ever-expanding,
stagnated, and shock-heated liner. The imploding liner for such a
bounce-free regime must have highly constrained profiles of density
$\rho$, pressure $p$, and flow speed $u$. 

This Letter demonstrates the existence of bounce-free spherical implosions by presenting a family
of imploding flow solutions that exactly satisfy the hydrodynamic equations and jump conditions at
$R_{\rm sh}(t)$ as well as the following constraint
\begin{equation}
\label{eq:bounce-free-constraint}
p(r,t) = {\rm const.}, \,\,\,\rho(r,t) = {\rm const.},\,\,\,{\rm and}\,\,
u(r,t)=0
\end{equation}
for $0<r<R_{\rm sh}(t)$ and $t> t_{\rm st}.$ Interestingly, such a
solution consistently describes cases of both an infinitesimally small
target (void) and a finite sized target kept still.  Indeed, the liner flow
solution with no distinct target satisfying
Eq.~(\ref{eq:bounce-free-constraint}) obeys all the first principle
equations even upon "inserting" the target into the center of symmetry,
while condition of the shocked liner pressure being constant ensures
that the target stays at rest. We notice that the solution we are aiming for is thus a generalization of a well known Noh's solution~\cite{Noh}. Indeed, Noh's solution gives the uniform pressure and density in the stagnated gas for the case, in which the initial converging shell, and therefore the the unshocked part of the liner as sketched in Figure~\ref{fig:sketch}, is perfectly cold, while the solution to be presented here allows temperature of the shell to be finite. Unlike Noh's case, it is then not possible to write a simple analytical formula describing liner evolution. It is still possible though to reduce the problem to a set of relatively simple ordinary differential equations, as we demonstrate later in this Letter.

The simplest model for spherically symmetric ideal hydrodynamic implosion is
the conservation laws for fluid mass, momentum, and entropy
\begin{eqnarray}
\rho_t + \frac{\Bigl(\rho u r^2\bigr)_r}{r^2} & = & 0,\label{eq:drhodt}\\
u_t + uu_r + \frac{p_r}{\rho} & = & 0,\label{eq:dudt}\\
\Bigl(p\rho^{-\gamma}\Bigr)_t + u\Bigl(p\rho^{-\gamma}\Bigr)_r & = & 0,
\label{eq:entropy}
\end{eqnarray} 
where the subscripts ``t'' and ``r' denote
differentiating over the time and radial variables, respectively.
These three equations need to be solved in the unshocked
region of the liner with the  jump condition at the shock front $R_{\rm
  sh}(t)$ connecting the infalling liner to the bounce-free
constraints, Eq.~(\ref{eq:bounce-free-constraint}).  To this end, it
is convenient to focus our consideration on the similarity solutions. A large class of such solutions and their application to ICF problems is considered in Ref.~\cite{Atzeni}.  In our case, its sub-class, the so-called spherical
quasi-simple waves, turns out to be sufficient. To relate these waves to those described in Ref.~\cite{Atzeni}, one should start from the more general form given by its Eqs. (6.182-6.183) and set $\alpha=1$ and $\kappa=0$. In the next paragraphs we follow \S\S 162-164 of
Ref.~\cite{courant-book-1976} to outline the basic features of this
solution family which we then utilize for the purpose of this Letter.

Spherical flow belongs to the class of quasi-simple waves if the
velocity, pressure, and density are constant on rays $\eta\equiv
t/r={\rm const}.$ An immediate consequence of this property is that the
flow is isentropic, i.e., $p\rho^{-\gamma}$ is not only conserved
for a given fluid element, as guaranteed by Eq.~(\ref{eq:entropy}),
but is also the same for all fluid elements. The pressure and density
can be then expressed in terms of the sound speed $c$ and with the
help of some algebraic manipulations, Eq.~({\ref{eq:drhodt}) and
  (\ref{eq:dudt}) reduce to
\begin{eqnarray}
D\eta U_\eta & = & \Bigl[ (1-U)^2 - 3 C^2 \Bigr] U,\label{eq:dUdt}\\
D\eta C_\eta & = & \Bigl[ (1-U)^2 - (\gamma-1) U(1-U) - C^2\Bigr] C,
\label{eq:dCdt}
\end{eqnarray}
where $D\equiv (1-U)^2 - C^2, \,\,\, U\equiv \eta u(\eta)$ and
$C\equiv \eta c(\eta).$ To write Eqs.~(\ref{eq:dUdt},\ref{eq:dCdt}),
pressure, density, velocity, the radial coordinate and time are normalized
to $p_{\rm st}, \rho_{\rm st}, \sqrt{p_{\rm st}/\rho_{\rm st}}$,
$r_{\rm st}$ and $r_{\rm st}/\sqrt{p_{\rm st}/\rho_{\rm st}}$, respectively.

It is instructive to consider possible scenarios on the $(U,C)$ plane.
To do so we divide Eq.~(\ref{eq:dCdt}) by Eq.~(\ref{eq:dUdt}) to
find
\begin{equation}
\label{eq:dCdU}
\frac{dC}{dU}
=\frac{(1-U)^2 - (\gamma-1) U(1-U) - C^2}{(1-U)^2 - 3C^2}
\frac{C}{U}.
\end{equation}
The full vector field of Eq.~(\ref{eq:dCdU}) is quite complex and
contains a number of different regimes (e.g., see Fig.~11 of Ch.~6
of Ref.~\cite{courant-book-1976}).  The $C(U)$ trajectories of interest (i.e., those
corresponding to the converging flow stopped by the shock
wave propagating outward) are located in the upper-left quadrant of the
plane, as plotted in Fig.~\ref{fig:CU-diagram}.  The arrows are in the direction of
growing $\eta,$ i.e., the direction of growing $t$ for fixed $r$ or
decreasing $r$ for fixed $t.$

The jump conditions along with
Eq.~(\ref{eq:bounce-free-constraint}) can be employed to find that $C_{\rm sh}$ and
$U_{\rm sh},$ values of $C$ and $U$ just before the shock front, must
obey
\begin{eqnarray}
(1-\mu^2) C_{\rm sh}^2 & = & (1-\mu^2)(1-U_{\rm sh})^2 + U_{\rm sh}(1-U_{\rm sh})
\label{eq:jump-condition} \\
U_{\rm sh} & = & (1-\mu^2)(1-c_{\rm st}^2/v_{\rm sh}^2),
\label{eq:shock-speed}
\end{eqnarray}
where $v_{\rm sh}$ is the shock speed and $\mu^2 \equiv
(\gamma-1)/(\gamma + 1).$ Eq.~(\ref{eq:jump-condition}) implies that the
shock transition takes place on an ellipse in the $(U,C)$
plane. Choosing a certain point on this ellipse then gives the shock
speed $v_{\rm sh}$ through Eq.~(\ref{eq:shock-speed}). Hence, the
desired regime is described by a curve uniquely defined by $v_{\rm
sh}$ that starts somewhere on the ellipse in the upper-left quadrant and
runs toward the origin of the $(U,C)$ plane, since this direction
corresponds to going to larger radii for a fixed $t.$ Importantly, $C$
and $U$ tend to zero together with $\eta,$ so that the sound speed
$c=\eta^{-1}C(\eta)$ and the fluid velocity $u=\eta^{-1}U(\eta)$ stay
finite.  For a given curve, $U$ and $C$ as functions of $\eta$ can be
found by integrating
Eqs.~(\ref{eq:jump-condition},\ref{eq:shock-speed}) from
$\eta=\eta_{\rm sh}$ to $\eta=0$ with initial conditions $U(\eta_{\rm
sh})=U_{\rm sh}$ and $C(\eta_{\rm sh}) = C_{\rm sh},$ where $\eta_{\rm
sh}=1/v_{\rm sh}.$ Once $U(\eta)$ and $C(\eta)$ are evaluated,
$u(\eta)$ and $c(\eta)$ can be recovered from definitions given after
Eq.~(\ref{eq:dCdt}), and $p(\eta)$ and $\rho(\eta)$ can be found
from the isentropicity condition. As a result, we obtain a solution
family parameterized by $v_{\rm sh}$ for bounce-free spherical imploding flows.

Any set of profiles $v_{v_{\rm sh}}(\eta)$ and $c_{v_{\rm sh}}(\eta)$
generated as described in the preceding paragraph explicitly answers
the question of what the liner profiles must be at and after stagnation
in order for the target to maintain constant
pressure and remain bounce-free. Next, we investigate the
evolution of the liner prior to stagnation to see whether
the liner profile found can be created in a practical device.  To do
so, we need to solve Eq.~(\ref{eq:dCdt},\ref{eq:dUdt}) backward in
time, i.e., extend the previously found solution from the upper-left
quadrant of Fig.~\ref{fig:CU-diagram} to the lower-right one. In
physical space, passing through the origin of
Fig.~\ref{fig:CU-diagram} corresponds to the liner reaching the center
of symmetry. Consequently, unlike the
previously considered post-stagnation solutions, this part of our analysis (for $t<0$) only qualitatively
applies to the physical case of a finite-size, preformed target. Nevertheless, 
even such a qualitative consideration provides substantial
insight into relevant liner parameters, as we demonstrate next.

As shown in Fig.~\ref{fig:CU-diagram}, a critical point exists on the
ellipse such that for $v_{\rm sh} > v_{\rm cr}$ the solution curve
ends up in the ``rest point'' ($U=0,\,\,\,C=-1$), whereas for $v_{\rm
sh} < v_{\rm cr}$ in the ``cavitation point'' ($U=1,\,\,C=0$).  The
former case corresponds to the situation in which initially the entire
space beyond the stagnation radius is occupied by a liner material at
rest with finite density and pressure. The latter case corresponds to
the situation in which initially there is a vacuum between the
infalling liner material and the target. This is the default regime
for a plasma liner driven MIF experiment. Despite behavior of all the
curves being qualitatively the same for $\eta>0,$ only those generated
from $v_{\rm sh} < v_{\rm cr}$ shall therefore be considered.  To show
the characteristic liner evolution in the cavitation regime we set
$v_{\rm sh}= 0.88 < v_{\rm cr}$ and integrate the fluid equations as
previously described to obtain $v_{v_{\rm sh}}(\eta), c_{v_{\rm
sh}}(\eta)$ and $p_{v_{\rm sh}}(\eta).$ Then, by inserting $t/r$ for
$\eta$ we generate a series of snapshots to create an
animation~\cite{animation} of pre- and post-stagnation phases of a bounce-free spherical implosion.

\begin{figure}[htbp]
\includegraphics[width=0.6\textwidth,keepaspectratio]{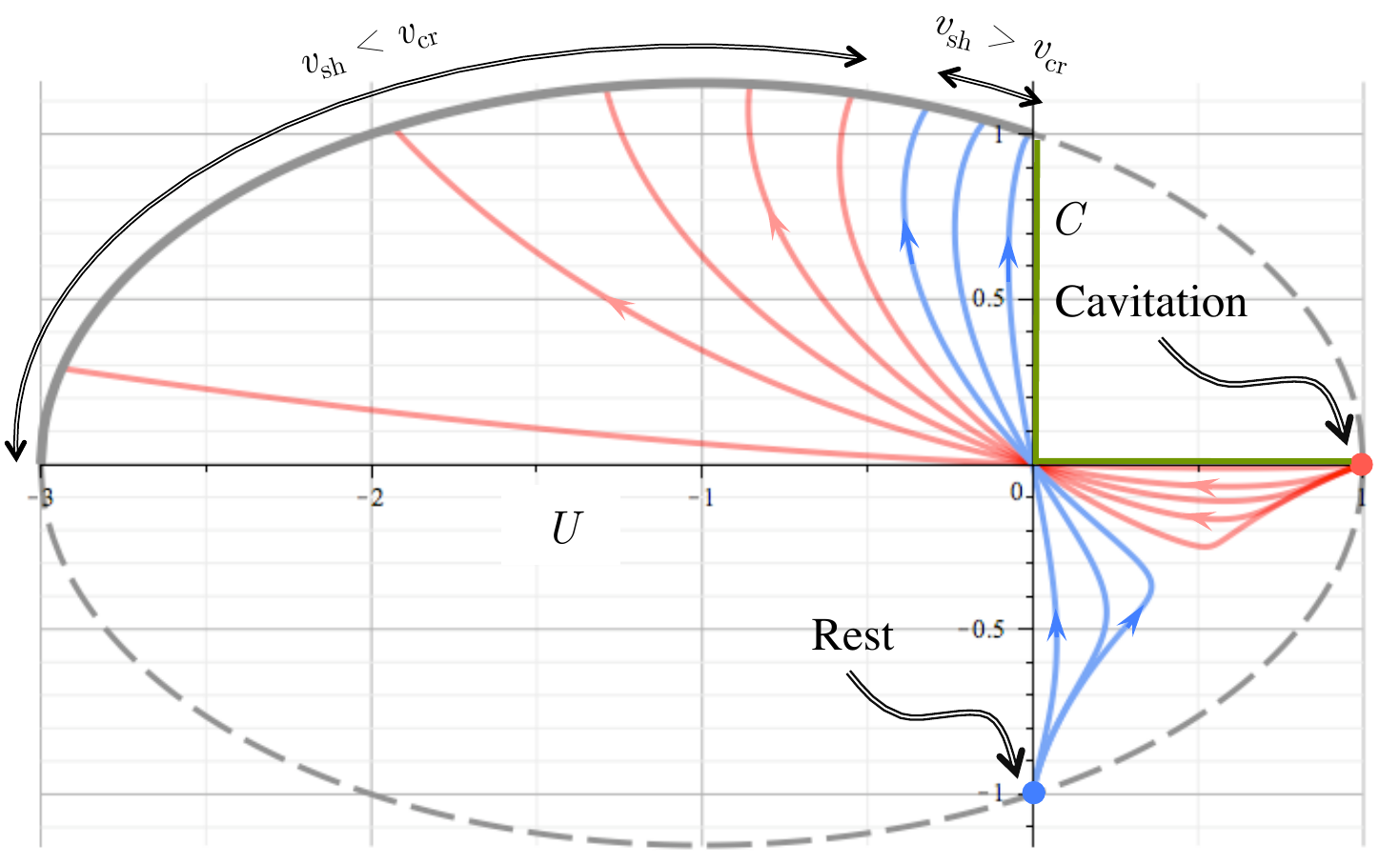}
\caption{Integral curves of Eq.~(\ref{eq:dCdU}) for $\gamma=5/3.$ The upper-left quadrant ($\eta>0$)
corresponds to the bounce-free implosion regime sketched in Fig.~\ref{fig:sketch}. The red branches in the 
lower-right quadrant ($\eta<0$) correspond to liner convergence in 
vacuum before reaching the origin. Trajectory in green corresponds to Noh's solution.}
\label{fig:CU-diagram}
\end{figure}

Having now proven the existence of bounce-free spherical hydrodynamic
implosion by explicitly constructing a family of self-similar
solutions to the spherically symmetric hydrodynamic equations, our
next objective is to elucidate the physics implications of this
regime. As previously mentioned, the physics advantages of bounce-free
implosion is probably most easily appreciated in the plasma liner
driven MIF concept. This is for two reasons.  The first is that
generally MIF schemes do not rely on an ignited burn wave because they
operate at ignition densities substantially lower than for ICF; the
desired high burn fraction is primarily achieved by maintaining the
compressed hot spot at thermonuclear burning condition for as long as
possible. A bounce-free implosion minimizes the decompressional
cooling of the target compared with a normal bounced implosion.  The
second is that by employing a standoff driver such as an array of
plasma guns to form and deliver the liner, one has substantial freedom
in shaping the liner profile (density, speed, and thermal pressure)
and the possibility of using a thick, as opposed to thin, liner.  It
is also of interest to note that because of the much smaller areal
density of MIF targets, compared with that of conventional ICF, the
mean free path of an alpha particle is much greater than the target
radius and neglecting fusion heating effect on the target pressure and
on the right side of Eq.~(\ref{eq:entropy}) is a useful approximation.
Next, we illustrate the potential energy gain by comparing the dwell time estimate of the
bounce-free implosion solution with an example of a conventional
implosion.

Discussion on which of the curves with $v_{\rm sh} < v_{\rm cr}$ would
work best for a realistic plasma liner MIF device is beyond the
scope of this Letter and will be presented elsewhere. In what
follows, we focus on the general effect of liner shaping toward bounce-free implosion to
demonstrate that employing a solution presented here can result in a
substantial dwell time improvement. To do so, it is convenient to take
the estimate of Ref.~\cite{parks-pop-2008}, where no special liner shaping
is assumed and a simple stationary flow solution is used to model the
liner profiles, as a reference point. Namely, we next consider a liner
of the newly found form whose energy and velocity, as well as the
relevant target parameters, are the same as in
Ref.~\cite{parks-pop-2008}. It is then reasonable to attribute the
difference between the dwell time predictions to that between the
liner shapes and, accordingly, whether or not the implosion results in target bounce.

The earlier estimate~\cite{parks-pop-2008} assumes that a target expands (bounces)
at a speed as fast as the shock front propagation for $t>t_{\rm st},$
and the fuel disassembly is considered to be complete once its radius
becomes $2r_{\rm st}$ as given in Eq.~(38) of
Ref.~\cite{parks-pop-2008}. The dwell time is therefore obtained by
computing $r_{\rm st}/v_{\rm sh},$ where $r_{\rm st}$ is 0.5~cm. The
shock speed is estimated to be half of the 100~km/s liner velocity and
is thus 50~km/s. As a result, the dwell time estimate is
100~ns. The target temperature is taken to be 10~keV, with a fuel
burn-up fraction of about 0.01, and the corresponding stagnation
pressure is 62.5 Mb. The total liner kinetic energy is 122 MJ
and the fusion energy gain factor $G$ is estimated to be about
2.6\%. We now proceed by conducting a similar estimate for the bounce-free liner
solution.

If it were possible to maintain the unshocked liner
flow with the prescribed profiles infinitely long, the target would never bounce and be
forever kept in a state of maximum compression. Practically, the
total liner energy is finite and we need to ``cut'' the liner profile at a
certain point to match the above-mentioned 122~MJ total liner energy constraint.  
Of course, such cutting of a fluid equation
solution results in a rarefaction wave spreading beyond the
outer boundary of the liner. However, until a moment $t_*,$ when the
shock wave hits the outer boundary of the liner, this rarefaction wave propagates
at the unshocked liner sound speed, which is generally much less than
the speed of the liner itself. It is therefore a good approximation that despite being
finite the liner evolves in space and time according to the exact
solution above for $t<t_*.$ At $t>t_*,$ the entire liner is
shocked and the rarefaction wave is no longer negligible, since the
shocked liner sound speed $c_{\rm st}$ is comparable to $v_{\rm sh}.$
Once the rarefaction wave hits the target, it finally starts
disassembling. Accordingly, the dwell time $\tau_{\rm dw}$ can be estimated by
\begin{equation}
\label{eq:dwell-time}
\tau_{\rm dw} = (t_* - t_{\rm st}) + t_{\rm rare},
\end{equation}
where $t_{\rm rare}$ is the time it takes the rarefaction wave to
travel from $R_*,$ the outer liner boundary position at $t=t_*,$ to
$r_{\rm st}.$ 
Note that we are neglecting the further finite time it
would take for the target to double its radius, which is the criterion for dwell time
used in Ref.~\cite{parks-pop-2008}. Thus,  our estimate, Eq.~(\ref{eq:dwell-time}), is a
conservative one.

To evaluate the first term on the right hand side of
Eq.~(\ref{eq:dwell-time}) we note that $t_*-t_{\rm st}=(R_*-r_{\rm
st})/v_{\rm sh},$ where $R_*$ can be calculated from liner energy
conservation since the hydrodynamic efficiency of the target
compression is low~\cite{parks-pop-2008,cassibry-pop-2009}. That is,
\begin{equation}
\label{eq:liner-energy}
\frac{4\pi(R_*^3 - r_{\rm st}^3)}{3}\frac{p_{\rm st}}{\gamma -1} = E_{\rm liner},
\end{equation}
where we have used the fact that the energy density of the shocked
liner is equal to $p_{\rm st}/(\gamma -1)$ and $E_{\rm liner}$ stands
for the total liner energy.  It should be noted that
Eq.~(\ref{eq:liner-energy}) gives $R_*\propto\ (\gamma-1)^{1/3}$,
making our estimate quite insensitive to uncertainties in $\gamma$.
To isolate the effect of bounce-free liner shaping, we insert into
Eq.~(\ref{eq:liner-energy}) the same values of 62.5 Mb, 0.5 cm, and 122
MJ for $p_{\rm st}, r_{\rm st},$ and $E_{\rm liner}$ as in
Ref.~\cite{parks-pop-2008}.  Eq.~(\ref{eq:liner-energy}) then gives
$R_* \approx 1.5 cm = 3 r_{\rm st}$ and, upon setting the liner and
shock front velocities 100 km/s and 50 km/s to match those of
Ref.~\cite{parks-pop-2008}, we find $t_* - t_{\rm st} \approx 200$~ns.

The second term on the right hand side of Eq.~(\ref{eq:dwell-time})
can be estimated by noticing that the rarefaction wave propagates at
$c_{\rm st}\approx v_{\rm sh},$ where, as in the preceding paragraph,
we set $v_{\rm sh}=50$~km/s. It then takes about $(R_*-r_{\rm
st})/v_{\rm sh} \approx 200$~ns for the rarefaction wave to reach the target and
Eq.~(\ref{eq:dwell-time}) gives $\tau_{\rm dw} \approx 400$~ns, which
is four times larger than its counterpart in the bounced implosion.  
Since the target parameters are chosen to
be the same, the fusion energy gain factor is four times larger as
well (even neglecting the target radius doubling time in the bounce-free case).

In conclusion, we have demonstrated that by shaping the profiles of an
imploding inertial pusher, the concept of bounce-free spherical
hydrodynamic implosion can be physically realized.  A family of
self-similar bounce-free solutions to the spherically symmetric ideal
hydrodynamic equations is explicitly found, along with a description
of their experimental accessibility.  For the specific application of
plasma liner driven MIF, we show that using bounce-free liner profiles
can substantially slow down the core plasma expansion.  As a result,
the dwell time, and therefore the fusion gain, can be noticeably
increased over an unshaped liner with equivalent kinetic
energy. Compared to the specific case treated in
Ref.~\cite{parks-pop-2008}, employing the bounce-free implosion regime
improves the dwell time and energy gain by at least a factor of
four. Furthermore, the newly found implosion regime supports the idea
of using deuterium-tritium fuel in the inner parts of the liner (as
suggested in Ref.~\cite{thio} and Ref.~\cite{Lindemuth}), which upon
becoming shocked will also burn, thus further increasing the
gain. Indeed, in such a regime, the post-shocked liner is at rest,
i.e., in contrast to previously considered schemes, the kinetic energy
of the original liner is entirely converted into internal energy. This
feature brings the temperature of the liner next to the target closer
to fusion relevant magnitudes, which may further improve the overall
efficiency of plasma liner driven MIF devices.

This work was supported by the Laboratory
Directed Research and Development (LDRD) program of LANL (Kagan and Tang) and the U.S. Department of Energy Office of Fusion
Energy Sciences under contract DE-AC52-06NA25396 (Hsu and Awe).

%\bibliographystyle{ieeetr}
%\bibliographystyle{prsty}
%\bibliography{/home/xtang/hg_home/dust/proposal/early-career/biblio}

\end{document}